\documentclass[11pt,a4paper]{article}
\pdfoutput=1
\usepackage{a4wide}
\usepackage{amsfonts}
\usepackage{amssymb}
\usepackage{amsmath}
\usepackage{graphicx}
\usepackage{booktabs}
\usepackage{array}
\usepackage{color}
\usepackage{ulem}
\usepackage[small,bf]{caption}
\usepackage{amsmath}
\usepackage{graphicx}
\usepackage{booktabs}
\usepackage{array}
\usepackage{color}
\usepackage{ulem}
\usepackage{caption}
\usepackage{cite}
\usepackage[usenames,dvipsnames]{xcolor}
\usepackage{subfigure}
\usepackage{verbatim}
\usepackage{dsfont}
\usepackage{bbm}
\usepackage{cite}
\usepackage[usenames,dvipsnames]{xcolor}
\usepackage{subfigure}
\usepackage{verbatim}
\usepackage{dsfont}
\usepackage{bbm}

\def\1{\mathbf{1}}
\def\3{\mathbf{3}}
\def\2{\mathbf{2}}

\allowdisplaybreaks[1]

\numberwithin{equation}{section}

\newcounter{mysubequation}[equation]

\DeclareMathOperator{\ci}{\text{i}}

\definecolor{pink}{rgb}{1.,.2,.8}

\begin{document}
\begin{titlepage}

\vspace*{-15mm}
\begin{flushright}
IFT-UAM/CSIC-17-021\\
FTUAM-17-5\\
NCTS-PH1701\\
TTP16-030
\end{flushright}
\vspace*{0.7cm}

\begin{center}
{ \bf\LARGE Neutrino Mass Sum Rules and\\[0.5em] Symmetries of the Mass Matrix}
\\[8mm]
Julia Gehrlein$^{\, a,b,c,}$ \footnote{E-mail: 
\texttt{julia.gehrlein@uam.es}},
Martin Spinrath$^{\, a,d,}$ \footnote{E-mail: \texttt{martin.spinrath@cts.nthu.edu.tw}}
\\[1mm]
\end{center}
\vspace*{0.50cm}
\centerline{$^{a}$ \it Institut f\"ur Theoretische Teilchenphysik, Karlsruhe Institute of Technology,}
\centerline{\it Engesserstra\ss{}e 7, D-76131 Karlsruhe, Germany}
\vspace*{0.2cm}
\centerline{$^{b}$ Now at \it
Departamento  de  F\'{\i}sica Te\'{o}rica,  Universidad  Aut\'{o}noma  de  Madrid,}
\centerline{\it Cantoblanco  E-28049  Madrid,  Spain}
\vspace*{0.2cm}
\centerline{$^{c}$ Now at \it Instituto  de  F\'{\i}sica  Te\'{o}rica  UAM/CSIC,}
\centerline{\it Calle Nicol\'{a}s Cabrera  13-15,  Cantoblanco  E-28049  Madrid,  Spain}
\vspace*{0.2cm}
\centerline{$^{d}$ Now at \it Physics Division, National Center for Theoretical Sciences,}
\centerline{\it National Tsing-Hua University, Hsinchu, 30013, Taiwan}
\vspace*{1.20cm}

\begin{abstract}
\noindent Neutrino mass sum rules have recently gained again more attention
as a powerful tool to discriminate and test various flavour models in the near
future. A related question which was not yet discussed fully satisfactorily was the origin
of these sum rules and if they are related to any residual or accidental
symmetry. We will address this open issue here systematically and find
previous statements confirmed. Namely, that the sum rules are not related
to any enhanced symmetry of the Lagrangian after family symmetry breaking
but that they are simply the result of a reduction of free parameters due to
skillful model building.
\end{abstract}
\end{titlepage}
\setcounter{footnote}{0}

\section{Introduction}

The origin of flavour is still an open issue in the Standard Model of particle physics (SM)
and most of its extensions. In the recent past a very popular approach is based
on (discrete) family symmetries which can easily explain the number of generations
and gives a very good leading order description of the neutrino mixing angles,
for recent reviews, see, e.g., \cite{flavour-reviews, King:2013eh}.

One particular prediction in plenty of these models is a so-called neutrino mass sum
rule, which relates the three complex neutrino eigenvalues with each other.
That means that the three masses can be described by two complex parameters only
\footnote{In principle, one could  imagine that the three masses are described by
one complex parameter only, but we are not aware of any such model with phenomenologically
viable predictions.}.
In Tab.~\ref{tab:overview_SR} we have given a list of all twelve sum rules known
in the literature. For recent detailed phenomenological studies
of the mass sum rules, see \cite{King:2013psa,Gehrlein:2015ena,Gehrlein:2016wlc,Agostini:2015dna}.
All mass sum rules can be parametrised according to~\cite{King:2013psa, Gehrlein:2015ena} as
\begin{align}
 s(m_1,m_2,&m_3,\phi_1,\phi_2;c_1,c_2,d,\Delta \chi_{13},\Delta \chi_{23}) \equiv \nonumber\\
 &c_1 \left(m_1 \text{e}^{-\text{i}\phi_{1}}\right)^d 
\text{e}^{\text{i}\Delta \chi_{13}}+
c_2 \left(m_2 \text{e}^{-\text{i}\phi_{2}}\right)^d 
\text{e}^{\text{i}\Delta \chi_{23}}
+m_3^d~ = 0 \;,
\label{eq:parametrisation_SR}
\end{align}
where $\phi_1$ and $\phi_2$  are the Majorana phases. The quantities $c_1$, $c_2$, $d$,
$\Delta \chi_{13}$, and $\Delta \chi_{23}$ are parameters which characterise the sum rule.

\begin{table}
\centering
\begin{tabular}{c c c c c c c}
\toprule
Sum rule & References & $c_1$&$c_2$&$d$&$\Delta \chi_{13}$&$\Delta \chi_{23}$ \\
\midrule
	1&~\cite{Barry:2010yk,Bazzocchi:2009da,Ding:2010pc,Ma:2005sha,Ma:2006wm,Honda:2008rs,Brahmachari:2008fn,Kang:2015xfa,SR1} &$1$&$1$&$1$&$\pi$&$\pi$\\
	2&~\cite{SR2} &$1$&$2$&$1$&$\pi$&$\pi$\\
	3&~\cite{Barry:2010yk,Ma:2005sha,Ma:2006wm,Honda:2008rs,Brahmachari:2008fn,Altarelli:2005yx,Chen:2009um,Chen:2009gy,Kang:2015xfa,SR3} &$1$&$2$&$1$&$\pi$&$0$\\
	4&~\cite{SR4} &$1/2$&$1/2$&$1$&$\pi$&$\pi$\\
	5&~\cite{SR5} &$\tfrac{2}{\sqrt{3}+1}$&$\tfrac{\sqrt{3}-1}{\sqrt{3}+1}$&$1$&$0$&$\pi$\\
	6&~\cite{Barry:2010yk,Bazzocchi:2009da,Ding:2010pc,Cooper:2012bd,SR6} &$1$&$1$&$-1$&$\pi$&$\pi$\\
	7&~\cite{Barry:2010yk,Altarelli:2005yx,Chen:2009um,Chen:2009gy,Altarelli:2009kr,SR7,Altarelli:2008bg} &$1$&$2$&$-1$&$\pi$&$0$\\
	8&~\cite{SR8} &$1$&$2$&$-1$&$0$&$\pi$\\
	9&~\cite{SR9} &$1$&$2$&$-1$&$\pi$&$\pi/2,3\pi/2$\\
	10&~\cite{SR10,Hirsch:2008rp} &$1$&$2$&$1/2$&$\pi,0,\pi/2$ &$0,\pi,\pi/2$ \\
	11&~\cite{SR11} &$1/3$&$1$&$1/2$&$\pi$&$0$\\
	12&~\cite{SR12} &$1/2$&$1/2$&$-1/2$&$\pi$&$\pi$\\
	\bottomrule
\end{tabular}
\caption{\label{tab:overview_SR}Summary table of the sum rules present in the literature.
The parameters $c_1, c_2, d, \Delta \chi_{13}$, and $\Delta \chi_{23}$ that characterise
them are defined in Eq.~\eqref{eq:parametrisation_SR}. In sum rules~9 and~10 two possible
signs appear which lead to two possible values of $\Delta \chi_{i3}$.
}
\end{table}

Since in these models we have less free parameters than observables one might wonder
if there is some underlying symmetry behind the mass sum rules. This is particularly
tempting since they emerge usually in models which have much more symmetry than the
SM including neutrino masses so that it could well be that the
full symmetry of the Lagrangian is actually not completely broken. 
A residual (or accidental) symmetry could then be responsible for the reduction
of free parameters in the mass matrix and result in a sum rule
\footnote{In this letter we only discuss mass sum rules. There also the well-known
mixing sum rules which originate from an additional breaking of the residual symmetries
in the charged lepton or neutrino sector. For a recent very detailed account of all
the possibilities, see, e.g., \cite{Girardi:2015rwa} and references therein.}.

On the other hand this intriguing idea is challenged by the fact that the same sum
rule emerges in models
with different symmetries. For instance, sum rule 6 from Tab.~\ref{tab:overview_SR}
with $1/\tilde{m}_1+ 1/\tilde{m}_2 - 1/\tilde{m}_3=0$, where $\tilde{m}_i$ are the three complex
neutrino masses,  is realised in
models with $A_4$ \cite{Barry:2010yk},  $S_4$ \cite{Bazzocchi:2009da}  and
$A_5$ \cite{Cooper:2012bd} symmetry. Furthermore, in \cite{Gehrlein:2015ena} we
have already tried to argue against some more fundamental principle behind the mass sum
rules by emphasizing that the only common feature of all this models is a reduction of parameters
in the neutrino mass matrix.
To be more precise the reduction of free parameters comes from an interplay of the choice of the
family symmetry, the choice of particle representations under this symmetry, and the way the family
symmetry is broken. Nevertheless, there is no general common recipe simply due to the fact
that there is no underlying symmetry argument as we will show in the following. In a sense
mass sum rules are  a mere result of skillful model building.
Note that this implies, that they can appear in direct, semi-direct and indirect models
(for this classification see, e.g.,~\cite{King:2013eh}) since they can never be
mapped to any subgroup of the family symmetry.

In this short letter we want to extend this previous discussion by giving more formal arguments
to show that the symmetry of the Lagrangian is not enhanced by a neutrino mass sum rule after
symmetry breaking and that the presence of a neutrino mass sum rule cannot be directly related
to any symmetry. This is different from the case of the mixing angle predictions where the Klein symmetry
of the neutrino mass matrix can be identified with some of the generators of the family symmetry,
for explanations and references, see \cite{flavour-reviews}.
Hence, the claim that neutrino mass sum rules have no deeper reason than
sophisticated model building is confirmed.

\section{Symmetries of Majorana mass terms}
\label{sec:M}

We will start by considering a Majorana mass term in the Lagrangian for the neutrinos
\begin{align}
\mathcal{L}_\nu\supset \frac{1}{2}\nu_L C^{-1}M_M \nu_L+\text{H.c.}~,
\end{align}
where $\nu_L$ contains the three left-handed neutrino fields, $C$ is the charge
conjugation matrix and $M_M$ is the symmetric, complex neutrino Majorana mass
matrix. The recent success of flavour model building with (discrete) symmetries is based on the
assumption that the Klein symmetry of the neutrino Majorana mass term is the remnant
of a bigger family symmetry, see, e.g., \cite{flavour-reviews}.
The generators of the residual symmetry $G$  of the mass matrix fulfill 
the symmetry condition
\begin{align}
S^T M_M S=M_M~,
\label{eq:res}
\end{align}
with unitary matrices $S  \in G$. Note that we restrict ourselves here to unitary
representations to keep the kinetic term canonical. The question is now, if
there are additional possibilities for $S$ if a mass sum rule is present which enhance the
symmetry $G$. In the following we will work in a basis, where $M_M$
is diagonal since an enhanced symmetry should be present in any basis and we are
only interested in the masses here. The advantage is, that in this basis the sum rule is
most simple and obvious. Note also that the characteristic polynomial of the mass matrix
is the same in the flavour and the mass basis.

We will begin our considerations with a more intuitive perturbative approach and
later discuss a general calculation.

In our setup $G$ could be maximally $U(3)$ and we can write
$S = \text{exp}(\text{i} \, \alpha_i  \, T_i)$, $i=1,\ldots, 9$ with the common eight generators 
$T_a$ ($a=1,\ldots, 8$) of $SU(3)$  and $T_9$ the generator of $U(1)$ \cite{ramond}.

If there is any continuous symmetry $G$ this would be expressed in
conditions on the generators $T_i$.
For a continuous Lie group we can expand eq.~\eqref{eq:res} in $\alpha_i$ to get up to
 $\mathcal{O}(\alpha_i^2)$
\begin{align}
 \text{i} \, \alpha_i \left( (T_i)^T M_M + M_M T_i \right) =0~ \text{ for }i=1,\ldots, 9 \,.
\label{eq:cons}
\end{align}
Using the explicit forms of the generators  and $M=\text{diag}(\tilde{m}_1,\tilde{m_2},\tilde{m}_3)$
we obtain the following conditions 
\begin{align}
\text{i} \, \alpha_1(\tilde{m}_1+ \tilde{m}_2)+\alpha_2(\tilde{m}_1-\tilde{m}_2)=0~,\\
\text{i} \, \alpha_4(\tilde{m}_1+ \tilde{m}_3)+\alpha_5(\tilde{m}_1-\tilde{m}_3)=0~,\\
\text{i} \, \alpha_6(\tilde{m}_2+ \tilde{m}_3)+\alpha_7(\tilde{m}_2-\tilde{m}_3)=0~,\\
2 \, \text{i} \, \tilde{m}_1 \left( \alpha_3+\frac{\alpha_8}{\sqrt{3}}+\alpha_9 \right)=0~,\\
2 \, \text{i} \, \tilde{m}_2 \left( -\alpha_3+\frac{\alpha_8}{\sqrt{3}}+\alpha_9 \right)=0~,\\
2 \, \text{i} \, \tilde{m}_3 \left( -\frac{2\alpha_8}{\sqrt{3}}+\alpha_9 \right)=0~.
\end{align}

Before we discuss the sum rule case we briefly want to review some well known cases.
In the case of three distinct, independent mass eigenvalues we get
from eq.~\eqref{eq:cons} $\alpha_i=0$ and we have no continuous symmetry apart from
the trivial one in this case.

If one of the masses vanishes while the two other are independent and non-zero we obtain
a relation between the diagonal generators which leads to a $U(1)$ symmetry for the massless
state. For example, if $\tilde{m}_3=0$  
we have the enhanced symmetry $ G = \text{exp}(\text{i} \, \alpha \, T)$
with $T=\text{Diag}(0,0,1)$ as expected.

In the case of two equal masses (and the third different and non-zero)
we obtain, for instance, if $\tilde{m}_2=\tilde{m}_3$ only   the $SO(2)$ generator
in the 1-2 block as again expected

Now for the interesting case, that $\tilde{m}_3$ is a function depending on the two other
masses we have explicitly checked that for all twelve sum rules in Tab.~\ref{tab:overview_SR}
all $\alpha_i=0$ as in the case for three distinct, independent masses. 

Another approach is to start now from a point in the parameter space which has a
well-known enhanced symmetry. If there would be an enhanced symmetry in the
case of a mass sum rule it should still be there after a small perturbation.
For instance, setting $\tilde{m}_3 \equiv \tilde{m}_2$
will fix $\tilde{m}_1$ for a given sum rule. But for this particular point we have a $SO(2)$
symmetry. If there is any non-trivial residual symmetry $G$ for a small perturbation of the
symmetric points it must be related to a small perturbation to the elements of $SO(2)$.

In the concrete case of  $\tilde{m}_3=\tilde{m}_1+2\tilde{m}_2$  we take
$\tilde{m}_3=\tilde{m}_2 (1+ \epsilon)$ with $\epsilon$ a small perturbation
from the enhanced symmetry point. The mass matrix is then
\begin{align}
M_M=\text{Diag}(- \tilde{m}_2 (1- \epsilon),  \tilde{m}_2 , \tilde{m}_2 (1+ \epsilon)) ~.
\label{eq:mdiag}
\end{align}
Now we can expand in all $\alpha_i$, $i \neq 6$,
in eq.~\eqref{eq:cons} and set the $\alpha_i$ to be of $\mathcal{O}(\epsilon)$.
The only solution to this equation is again $\alpha_i=0$,
$i=1,\ldots 9$. This is also true in the case of other sum rules with different coefficients
as it can be easily understood from considering only the 2-3 block of $M_M$ which
exhibits a $SO(2)$ symmetry for $\epsilon=0$. In the case of
$\epsilon\neq 0$ the eigenvalues are different and we find no symmetry anymore.

We have also expanded around the other symmetry points for all sum rules
and around the well-known non-trivial $\mathbb{Z}_2^3$ symmetry
of the Majorana mass matrix (which corresponds
to expanding $\alpha_3$, $\alpha_8$ and $\alpha_9$ around $\pi$) but never
found any non-trivial solution for the $\alpha_i$.

Up to now we have only considered continuous symmetries where we can expand in
small $\alpha_i$ around the the elements of $\mathbb{Z}_2^3$ and concluded that
the presence of a sum rule does not enhance the
symmetry of the mass matrix. One might wonder if the residual symmetry we
are looking for is not anywhere near these points which would be surprising
but cannot be completely ruled out at this point.

To rule this out as well we also did the general calculation for an arbitrary
$S \in U(3)$ which is nevertheless tedious and not very insightful compared
to the perturbative approach. We will discuss here how to do this for sum rule 1,
cf.\ Tab.~\ref{tab:overview_SR}, as an example. For the other sum rules similar calculations
give the same result as we have checked. An element $S \in U(3)$ can be written as
$S = P_1 U_{23} U_{13} U_{12}$, for notation and conventions, see Appendix~A
of \cite{Gehrlein:2016wlc}. Since $S$ is unitary we can rewrite eq.~\eqref{eq:res}
\begin{equation}
 M_M S = S^* M_M \;. \label{eq:res2}
\end{equation}
The 1-1 element of this equation reads 
\begin{equation}
m_1 \cos \theta_{12} \cos \theta_{13} \left( \text{e}^{2 \ci \omega_1} -1 \right) = 0\;,
\end{equation}
which has four possible solutions. Let us discuss first $\theta_{13} = \pi/2$
(note that in our conventions $\theta_{ij} \in [0,\pi]$).
From the 1-3 element of eq.~\eqref{eq:res2} we find
that
\begin{equation}
\delta_{13} = \omega_1 \text{ and } m_2 = 2 \, m_1 \cos (  \phi_2 - \phi_1 )  \;,
\end{equation}
which is in general not satisfied
and we exclude the solution with $\theta_{13} = \pi/2$. 
For the same reason we have to discard
$\theta_{12} = \pi/2$. From the remaining two solutions
$\omega_2 = 0$ or $\pi$ it is sufficient to discuss $\omega_2 = 0$.
At this point they are related by a global sign.

From the 1-2 element of eq.~\eqref{eq:res2} we derive
\begin{equation}
 \left( m_1 \text{e}^{ \ci \phi_2} - m_2 \text{e}^{ \ci (\phi_1 + 2 \delta_{12} )}  \right) \cos \theta_{13} \sin \theta_{12} = 0 \;.
\end{equation}
As we have discussed above we have to discard the solution
$\theta_{13} = \pi/2$ and the only two remaining solutions are
$\theta_{12} = 0$ or $\pi$. For simplicity, we will only discuss here
$\theta_{12} = 0$. From the 1-3 element of eq.~\eqref{eq:res2}
we then find that
\begin{equation}
 \left( m_1 (\text{e}^{2  \ci \delta_{13} } -1) - m_2 \text{e}^{ \ci (\phi_1 - \phi_2 + 2 \delta_{13} )}  \right)  \sin \theta_{13} = 0 \;.
\end{equation}
Again $\theta_{13} = 0$ or $\pi$ and we discuss only  $\theta_{13} = 0$.
From the 3-2 element of eq.~\eqref{eq:res2}
we find
\begin{equation}
 \left( m_1  \text{e}^{\ci ( \phi_2 + 2 \delta_{23} + 2 \omega_3) } + m_2 \text{e}^{\ci ( \phi_1 + 2 \delta_{23} + 2 \omega_3) }   - m_2 \text{e}^{ \ci \phi_1 }  \right) \sin \theta_{23} = 0 \;.
\end{equation}
And hence $\theta_{23} = 0$ or $\pi$.

Now we know that $S$ has to be diagonal and it is trivial to see
that the remaining phases have to take trivial values.
So we have shown without resorting to any expansion that $S$ can be only
an element of $\mathbb{Z}_2^3$,
i.e.\ it must be a diagonal matrix with $\pm 1$ on the diagonal.
For the other sum rules we have checked with similar calculations
that $S$ can be only an element of $\mathbb{Z}_2^3$.

Hence, we conclude that there is no particular residual (continuous or discrete)
symmetry in the case of a
neutrino mass sum rule. This is actually not surprising. Apart from the ubiquitous
symmetric points where masses are equal or vanish even in the case of neutrino mass sum
rules the three neutrino masses are different which is known to have no other symmetry than
the $\mathbb{Z}_2^3$ (physically the Klein symmetry  $\mathbb{Z}_2^2$ after absorbing
an unphysical sign corresponding to one of the $\mathbb{Z}_2$ factors).

\section{Symmetries of Dirac mass terms}
\label{sec:D}

We turn now to the case of Dirac mass matrices which is nevertheless only
realised in one of the known flavour models exhibiting a mass sum rule 
\cite{Ding:2013eca} in the literature (this sum rule can also be realised in models with Majorana neutrinos \cite{Lindner:2010wr}).

One has to be careful since in the mass sum rule for Dirac neutrinos the Majorana phases are unphysical. Nevertheless, these sum rules still lead to  one of the major predictions of mass sum rules, the lower bound on the lightest mass.

We will show that also in the case of  Dirac neutrinos a mass sum rule does not
lead to any particular symmetry of the mass matrix.

A Dirac mass term reads
\begin{align}
\mathcal{L}_\nu\supset -\bar{\nu}_L M_D \nu_R
\end{align}
with the left- and right-handed neutrino fields $\nu_L$ and  $\nu_R$.

For a Dirac mass matrix $M_D$ the relation for an enhanced residual
or accidental symmetry $H$ is 
\begin{align} \label{eq:res_Dirac}
R^\dagger M^\dagger_D M_D^{\phantom{\dagger}} R= M^\dagger_D M_D^{\phantom{\dagger}}
\end{align}
with unitary matrices $R \in H$.
Again  $H$ can be maximally $U(3)$ and we set
$R=\text{exp}(\text{i} \, \beta_i \, T_i)$, $i=1,\ldots, 9$.
Now we have to find the solution for
 \begin{align}
 \text{i} \, (-\beta_i(T_i)^\dagger M_D^\dagger M_D^{\phantom{\dagger}} +M^\dagger_D M_D^{\phantom{\dagger}}  \beta_i T_i) =0~.
 \label{eq:dcon}
 \end{align}
In the case of three distinct eigenvalues we obtain
that  $\beta_3$, $\beta_8$ and $\beta_9$ are undetermined which leads to
a $U(1)^3$ symmetry corresponding to individual neutrino
flavour numbers.
This also does not change for a vanishing mass this time.

For two equal masses we get additionally that $\beta_1$ and $\beta_2\neq0$
are undetermined for $\tilde{m}_1=\tilde{m}_2$ which corresponds to $U(2) \times U(1)$ rotations.

In the case of a sum rule we find again that the symmetry group is not enhanced
for phenomenologically relevant parameter points. This is true also in the vicinity of
symmetry points as discussed above. And for definiteness we have also checked here
our statement for an arbitrary element of $U(3)$.
So again neutrino mass sum rules do not
lead to any particular residual symmetry of the Lagrangian also in the case
of Dirac neutrino masses.

\section{Summary and conclusions}
\label{sec:end}

Neutrino mass sum rules are a powerful way to test more than 60 flavour
models. Although the phenomenology of these models has been studied already
in great detail \cite{King:2013psa, Gehrlein:2015ena, Gehrlein:2016wlc, Agostini:2015dna}
the exact origin of the neutrino mass sum rules had
not been addressed  systematically yet.

Since they usually appear in the context of non-Abelian discrete family symmetries
it is tempting to think of them in the same framework and try to connect the neutrino
mass sum rules to any residual symmetry of the Lagrangian (the neutrino mass terms).
We have demonstrated here that this is not the case. From the viewpoint of residual
symmetries there is no difference between a mass matrix which fulfills a neutrino mass sum rule
and a mass matrix which does not.
n that sense mass sum rules present themselves as a model building artifact
found in many flavour models which have no one-to-one
mapping to any definite property of the flavour model in the unbroken
phase. Despite that they still offer robust predictions for the
neutrino mass ordering and scale, for instance, which can be
tested in the future.

Starting from the symmetry conditions, eqs.~\eqref{eq:res} and \eqref{eq:res_Dirac},
we have provided perturbative arguments and shown an explicit (non-perturbative)
example calculation which prove
our statement. We have checked that the given statements and calculations indeed extend
to all sum rules and 
we conclude that a neutrino mass sum rule is not related to an enhanced or
particular residual symmetry
of the Lagrangian as long as all the masses are distinct. This is in complete agreement with
the widely used claim that non-Abelian family symmetries cannot determine
the neutrino masses.
In fact, our proof is equally valid for the case without any mass
sum rule (three completely independent masses). To our knowledge, this is the first
formal proof of this widely held conviction in the literature.

These considerations clarify and settle an open question in the literature and prove that
neutrino mass sum rules are simply related to some minimal breaking of the symmetries
in flavour models in the sense that only the minimal set of parameters is introduced
in the neutrino mass matrix to allow for three non-vanishing eigenvalues.

\section{Acknowledgements}

We would like to thank Alexander Merle for useful discussions on the manuscript. JG acknowledges partial support  from the European Union's Horizon 2020 research and innovation programme under the Marie Sklodowska-Curie grant agreement No 674896. MS would like to thank Academia Sinica for kind hospitality during some stages
of this project and acknowledges partial support by BMBF under contract no.\ 05H12VKF.

\end{document}